\documentclass[a4paper,12pt]{article}

\date {}
\usepackage{amssymb,amsfonts,amsmath,epsfig,float,graphicx,subfigure}
\usepackage{amsmath}

\newcommand{\be}{\begin{eqnarray}}
\newcommand{\ee}{\end{eqnarray}}
\newcommand{\Tr}{\,{\rm Tr}\,}

\def\Io{{\mathbb I}}

\newcommand{\SU}{$\mathfrak{s}\mathfrak{u}$}

\title{On the BCH formula of Rezek and Kosloff}

\author{Jan Naudts and Winny O'Kelly de Galway\\
\small Departement Fysica, Universiteit Antwerpen,\\
\small Universiteitsplein 1, 2610 Antwerpen, Belgium
}

\begin{document}
\maketitle

\begin{abstract}
The BCH formula of Rezek and Kosloff is a convenient tool to handle
a family of density matrices, which occurs in the study of quantum heat engines.
We prove the formula using a known argument from Lie theory. 
\end{abstract}

\section{Introduction}

The Hamiltonian
\be
H=\frac 12\hbar\omega(a^\dagger a+aa^\dagger)
\ee
of the quantum harmonic oscillator belongs to the Lie algebra \SU(1,1) with generators
\be
S_1&=&\frac 14((a^\dagger)^2+a^2),\crcr
S_2&=&\frac i4((a^\dagger)^2-a^2),\crcr
S_3&=&\frac 14(a^\dagger a+aa^\dagger).
\ee
As a consequence, it is possible to write down simplified Baker-Campbell-Haussdorf (BCH) relations
\cite {WN63,WN64}.
These have been used to study the quantum harmonic oscillator in a time-dependent external field
\cite {LR69,DPB92,YKUGP94,SDY00,DL08,KN10,NOK10}.
The topic of the present paper is a new BCH relation, introduced
recently by Rezek and Kosloff \cite {RK06}. They consider the family of density matrices
\be
\rho=\frac 1{Z(\beta,\gamma)}e^{\gamma a^2}e^{-\beta H}e^{\overline\gamma (a^\dagger)^2}
\label{densop}
\ee
with real $\beta$ and complex $\gamma$, and with
\be
Z(\beta,\gamma)=\Tr e^{\gamma a^2}e^{-\beta H}e^{\overline\gamma (a^\dagger)^2}.
\ee
Because all operators appearing in (\ref {densop}) belong to the Lie algebra it is clear from the
general Baker-Campbell-Haussdorf relation that it must be possible to write
\be
e^{\gamma a^2}e^{-\beta H}e^{\overline\gamma (a^\dagger)^2}=e^{\chi a^2-\xi H+\overline\chi (a^\dagger)^2}
\label{explic}
\ee
The explicit expression of the coefficients $\chi$ and $\xi$ as a function of $\beta$ and $\gamma$
is found in the Appendix of \cite {RK06}. The functions were derived \cite {RY10} using the
algebraic manipulation software Mathematica.

Note that the special case of (\ref {explic}) with $\xi=0$
appeared in the physics literature before (see Example I of Section II of \cite {GR74}; see also \cite {FNS84,TDR85}).
In the present paper the relation
(\ref {explic}) is derived using the argument of \cite {GR74}.

The relation (\ref {explic}) is of interest in its own. But it is also very useful in the study of
quasi-stationary processes \cite {RK06,SHRK09}. Indeed, from (\ref {explic}) it is clear that
the density matrix $\rho$ describes a system in thermal equilibrium at inverse temperature $\xi$.
On the other hand, the expression (\ref {densop}) is more convenient for practical calculations.

We derive the BCH formula in the next section.
In Section 3 follows a similar BCH formula valid for \SU(2).
In the final section follows a short discussion.

\section{The identity}
\label {section:identity}

The the l.h.s.~of (\ref {explic}) can be written in terms of the generators of the Lie algebra \SU(1,1) as
\be
e^{2\gamma(S_1+iS_2)}e^{-2\beta\hbar\omega S_3}
e^{2\overline\gamma (S_1-iS_2)}.
\label {rhosu11}
\ee
These generators satisfy the commutation relations
\be
[S_1,S_2]&=&iS_3,\cr
[S_2,S_3]&=&-iS_1,\cr
[S_3,S_1]&=&-iS_2.
\ee
Introduce \SU(2) generators $T_1=-iS_1$, $T_2=iS_2$, and $T_3=S_3$.
Then (\ref {rhosu11}) becomes
\be
X
\equiv e^{2\gamma(iT_1+T_2)}e^{-2\beta\hbar\omega T_3}
e^{2\overline\gamma (iT_1-T_2)}.
\label {rhosu2}
\ee
The relation (\ref {explic}) does not depend on the choice of
the representation of the \SU(2) algebra. Therefore,
we may change it. A favourable choice is that
of the Pauli spin matrices $\sigma_\alpha=2T_\alpha$.
Using that $(\sigma_1\pm i\sigma_2)^2=0$ and $\sigma_\alpha^2=\Io$
the calculation becomes very easy. One obtains
\be
X
&=&e^{i\gamma(\sigma_1-i\sigma_2)}e^{-\beta\hbar\omega \sigma_3}
e^{i\overline\gamma (\sigma_1+i\sigma_2)}\cr
&=&(\Io+i\gamma(\sigma_1-i\sigma_2))(\cosh(\hbar\omega)-\sigma_3\sinh(\hbar\omega))
(\Io+i\overline\gamma (\sigma_1+i\sigma_2))\cr
&=&e^{-\beta\hbar\omega \sigma_3}
-2\kappa|\gamma|^2+i\kappa(\gamma+\overline\gamma)\sigma_1+\kappa(\gamma-\overline\gamma)\sigma_2
+2\kappa|\gamma|^2\sigma_3,\cr
& &
\label{identity_temp}
\ee
with $\kappa=e^{-\beta\hbar\omega}$ as before.
On the other hand is
\be
\exp\left(\chi a^2-\xi H+\overline\chi (a^\dagger)^2\right)
&=&\exp\left(2\chi(S_1+iS_2)-2\xi\hbar\omega S_3+2\overline\chi(S_1-iS_2)\right)\cr
&=&\exp\left(2\chi(iT_1+T_2)-2\xi\hbar\omega T_3+2\overline\chi(iT_1-T_2)\right).\cr
& &
\ee
In the Pauli spin representation this becomes $e^{Y}$ with
\be
Y=i(\chi+\overline\chi)\sigma_1+(\chi-\overline\chi)\sigma_2-\xi\hbar\omega\sigma_3.
\ee
Because the Pauli matrices anti-commute and their squares equal $\Io$ there follows that
\be
Y^2=\lambda^2\Io\quad\mbox{ with }\quad \lambda=\sqrt{\xi^2(\hbar\omega)^2-4|\chi|^2}.
\label {lambdadef}
\ee
Hence one obtains
\be
e^{Y}=\cosh(\lambda)+\frac 1\lambda \sinh(\lambda)Y.
\ee
Comparison with (\ref {identity_temp}) gives the 4 conditions
\be
\cosh(\lambda)&=&\alpha+\kappa,
\label {eq1}\\
\frac 1\lambda\sinh(\lambda)(\chi+\overline\chi)&=&\kappa(\gamma+\overline\gamma),
\label {eq2}\\
\frac 1\lambda\sinh(\lambda)(\chi-\overline\chi)&=&\kappa(\gamma-\overline\gamma),
\label {eq3}\\
\frac 1\lambda\sinh(\lambda)\xi\hbar\omega&=&\alpha,
\label {eq4}
\ee
with
\be
\alpha&=&\sinh(\beta\hbar\omega)-2\kappa|\gamma|^2\cr
&=&\frac 1{2\kappa}\left[1-\kappa^2-4\kappa|\gamma|^2\right].
\label {alphadef}
\ee
The solution of these equations is
\be
\xi&=&\frac \alpha{\hbar\omega}\,\frac \lambda{\sinh(\lambda)},
\label {xires}\\
\chi&=&\kappa\frac \lambda{\sinh(\lambda)}\gamma.
\label {soleq}
\ee
with
\be
\sinh(\lambda)=\sqrt{\alpha^2-4\kappa^2|\gamma|^2}
\ee
These results coincide with those found in the Appendix of \cite {RK06}.

Note that the expressions for $\xi$ and $\chi$ can be inverted easily.
Given $\xi$ and $\chi$ one obtains $\lambda$ from (\ref {lambdadef}).
Then $\alpha$ follows by inverting (\ref {xires}). This gives
\be
\alpha=\hbar\omega\xi\frac {\sinh(\lambda)}\lambda.
\ee
Next $\beta$ is obtained from (\ref {eq1})
\be
\kappa=\cosh(\lambda)-\alpha.
\ee
Finally, $\gamma$ follows from (\ref {soleq})
\be
\gamma=\frac {\sinh(\lambda)}{\kappa\lambda}\chi.
\ee

\section{An example with SU(2) symmetry}

Formulas similar to (\ref {explic}) can be derived for other symmetry groups than SU(1,1).
For instance, in the case of SU(2) one has
\be
e^{\gamma\sigma_+}e^{-\beta\sigma_z}e^{\overline\gamma\sigma_-}
=\exp\left(\chi\sigma_+-\xi\sigma_z+\overline\chi\sigma_-\right)
\label {su2form}
\ee
with $\sigma_\pm=\frac 12(\sigma_x\pm i\sigma_y)$.
Using $\sigma_\pm^2=0$, $\sigma_z^2=\Io$, $\sigma_\pm\sigma_z=\mp\sigma_\pm$,
 and $\sigma_+\sigma_-=\frac 12(1+\sigma_z)$ the l.h.s.~becomes
\be
\mbox{l.h.s.}
&=&\left(1+\gamma\sigma_+\right)
\left(\cosh(\beta)-\sinh(\beta)\sigma_z\right)
\left(1+\overline\gamma\sigma_-\right)\cr
&=&\cosh(\beta)+\frac 12|\gamma|^2e^\beta
+e^\beta(\gamma\sigma_++\overline\gamma\sigma_-)
-(\sinh(\beta)-\frac 12e^\beta|\gamma|^2)\sigma_z.\cr
& &
\ee
The r.h.s.~of (\ref {su2form}) is evaluated using
\be
\left(\chi\sigma_+-\xi\sigma_z+\overline\chi\sigma_-\right)^2=\lambda^2\Io,
\ee
with $\lambda=\sqrt{\xi^2+|\chi|^2}$.
One finds
\be
\mbox{r.h.s.}
&=&\cosh(\lambda)
+\frac 1{\lambda} \sinh(\lambda)
 \left(\chi\sigma_+-\xi\sigma_z+\overline\chi\sigma_-\right).
\ee
Equating both expressions yields the set of equations
\be
\cosh(\beta)+\frac 12e^\beta|\gamma|^2&=&\cosh(\lambda),
\label {su2_1}\\
-\sinh(\beta)+\frac 12e^\beta|\gamma|^2&=&-\frac 1\lambda \sinh(\lambda)\xi,
\label {su2_2}\\
\gamma e^\beta&=&\frac 1\lambda \sinh(\lambda)\chi.
\label {su2_3}
\ee
Given $\xi$ and $\chi$, the value of $\lambda$ can be obtained from its definition.
The solution then reads
\be
e^\beta&=&\cosh(\lambda)+\frac 1\lambda \sinh(\lambda)\xi\cr
\gamma&=&\frac {\frac 1\lambda \sinh(\lambda)}{\cosh(\lambda)+\frac 1\lambda \sinh(\lambda)\xi}\chi.
\ee
Conversely, given $\beta$ and $\gamma$ one obtains $\lambda$ from (\ref {su2_1}).
Then $\xi$ and $\chi$ follow from (\ref {su2_2}) and (\ref {su2_3}), respectively.

\section{Discussion}

The BCH relation of Rezek and Kosloff
is somewhat special because it is written in a form suited
for application to density matrices.
Similar results found in the literature
\cite {WN63,WN64,LR69,DPB92,YKUGP94,SDY00,DL08,KN10,NOK10}
aim at the calculation of time evolution operators and refer to similarity transformations,
this is, to expressions of the form $e^ABe^{-A}$.
But the l.h.s.~of (\ref {explic}) is not a similarity transformation. This is precisely the reason why this
BCH relation is of interest! The change of the spectrum implies that the average energy
$\langle H\rangle=\Tr\rho H$ will depend on the value of the parameter $\gamma$.
This dependence is essential in the context of heat engines.

\section*{}

\end{document}